\documentclass[onecolumn]{aastex63}
\usepackage{rotating} 
\usepackage{amsmath,amssymb,bbold,mathrsfs}
\usepackage{subfigure}
\usepackage{bm}

\makeatletter
\let\frontmatter@title@above=\relax
\makeatother

\shorttitle{Magnetic stratification inference using the whole CLASP2/2.1 spectral window}
\shortauthors{Afonso Delgado et al.}

\graphicspath{{./}{figures/}}

\begin{document}

\title{Determining the magnetic field of active region plages \\
using the whole CLASP2/2.1 spectral window}

\email{dadelgado@ucar.edu}

\author[0000-0002-4163-4347]{David Afonso Delgado}
\altaffiliation{Present Address: High Altitude Observatory (NCAR), Boulder, CO, USA.}
\affil{Instituto de Astrof\'{\i}sica de Canarias, E-38205 La Laguna, Tenerife, Spain}
\affil{Departamento de Astrof\'\i sica, Universidad de La Laguna, E-38206 La Laguna, 
Tenerife, Spain}

\author[0000-0003-1465-5692]{Tanaus\'u\ del Pino Alem\'an}
\affil{Instituto de Astrof\'{\i}sica de Canarias, E-38205 La Laguna, Tenerife, Spain}
\affil{Departamento de Astrof\'\i sica, Universidad de La Laguna, E-38206 La Laguna, 
Tenerife, Spain}

\author[0000-0001-5131-4139]{Javier\ Trujillo Bueno}
\altaffiliation{Affiliate Scientist of the High Altitude Observatory (NCAR), Boulder, CO, USA.}
\affil{Instituto de Astrof\'{\i}sica de Canarias, E-38205 La Laguna, Tenerife, Spain}
\affil{Departamento de Astrof\'\i sica, Universidad de La Laguna, E-38206 La Laguna,
 Tenerife, Spain}
\affil{Consejo Superior de Investigaciones Cient\'{\i}ficas, Spain}

\author[0000-0001-8830-0769]{Ryohko\ Ishikawa}
\affiliation{National Astronomical Observatory of Japan, 2-21-1 Osawa, Mitaka, Tokyo 
181-8588, Japan}

\author[0000-0001-9095-9685]{Ernest\ Alsina Ballester}
\affil{Instituto de Astrof\'{\i}sica de Canarias, E-38205 La Laguna, Tenerife, Spain}
\affil{Departamento de Astrof\'\i sica, Universidad de La Laguna, E-38206 La Laguna, 
Tenerife, Spain}

\author[0000-0002-9921-7757]{David E. McKenzie}
\affiliation{NASA Marshall Space Flight Center, Huntsville, AL 35812, USA}

\author[0000-0002-8775-0132]{Luca Belluzzi}
\affiliation{Istituto ricerche solari Aldo e Cele Daccò (IRSOL), Faculty of Informatics,
 Università della Svizzera italiana (USI), CH-6605 Locarno, Switzerland}
\affiliation{Institut für Sonnenphysik (KIS), D-79110, Freiburg i. Br., Germany}
\affiliation{Euler Institute, Universit\`a della Svizzera italiana (USI), 
CH-6900 Lugano, Switzerland}

\author{Christian Bethge}
\affiliation{Cooperative Institute for Research in Environmental Sciences,
University of Colorado Boulder, Boulder, CO 80305, USA}

\author[0000-0003-1057-7113]{Ken Kobayashi}
\affiliation{NASA Marshall Space Flight Center, Huntsville, AL 35812, USA}

\author[0000-0003-3765-1774]{Takenori J. Okamoto}
\affiliation{National Astronomical Observatory of Japan, 2-21-1 Osawa, Mitaka, 
Tokyo 181-8588, Japan}

\author[0000-0002-3770-009X]{Laurel A. Rachmeler}
\affiliation{National Oceanic and Atmospheric Administration, \\ 
National Centers for Environmental Information, Boulder, CO 80305, USA}

\author[0000-0003-3034-8406]{Donguk Song}
\affiliation{Korea Astronomy and Space Science Institute, 776 Daedeok-daero, 
Yuseong-gu, Daejeon 34055, Republic of Korea}
\affiliation{National Astronomical Observatory of Japan, 2-21-1 Osawa, Mitaka, Tokyo 181-8588, Japan}

\author[0000-0002-8292-2636]{Ji\v{r}\'i  \v{S}t\v{e}p\'an}
\affil{Astronomical Institute, Academy of Sciences of the Czech Republic, 25165 Ondrejov, Czech Republic}

\author[0000-0002-8370-952X]{Bart de Pontieu}
\affil{Lockheed Martin Solar \& Astrophysics Laboratory, Palo Alto, CA 94304, USA}
\affil{Rosseland Centre for Solar Physics, University of Oslo, P.O. Box 1029 Blindern, N-0315 Oslo, Norway}
\affil{Institute of Theoretical Astrophysics, University of Oslo, P.O. Box 1029 Blindern, N-0315 Oslo, Norway}

\author[0000-0002-4691-1729]{Adam R. Kobelski}
\affiliation{NASA Marshall Space Flight Center, Huntsville, AL 35812, USA}

\author[0000-0002-7219-1526]{Genevieve D. Vigil}
\affiliation{NASA Marshall Space Flight Center, Huntsville, AL 35812, USA}

\author[0000-0003-0972-7022]{Fr\'ed\'eric Auch\`ere}
\affiliation{Institut d'Astrophysique Spatiale, 91405 Orsay Cedex, France}

\author[0000-0002-2093-085X]{Ryouhei Kano}
\affiliation{National Astronomical Observatory of Japan, 2-21-1 Osawa, Mitaka, Tokyo 
181-8588, Japan}

\author[0000-0002-5608-531X]{Amy Winebarger}
\affiliation{NASA Marshall Space Flight Center, Huntsville, AL 35812, USA}

\begin{abstract}

The Chromospheric LAyer SpectroPolarimeter missions, CLASP2 and CLASP2.1, demonstrated
that the near-UV spectral region between 279.30 and 280.68~nm 
is suitable for studying the magnetism of the solar chromosphere. 
In particular, the spectropolarimetric observations in the  \ion{Mg}{2} h and k resonant doublet,
\ion{Mn}{1} 279.91 and 280.19~nm resonant lines, 
and \ion{Fe}{2} 279.79 and 280.66~nm lines acquired by these suborbital space experiments
have been proven useful for inferring the magnetic field stratification in the solar chromosphere.
However, several lines of the CLASP2/2.1 spectral region 
with significant circular polarization signals had remained unexplored.
After identifying two \ion{Ni}{1} (279.95 and 280.59~nm), one \ion{Mn}{2} (280.62~nm), 
and one \ion{Fe}{1} (280.53~nm) lines, here we apply the Weak Field Approximation (WFA) to the spectropolarimetric 
observations of active region plages by CLASP2 and CLASP2.1.
By comparing the results with previous studies, 
we are able to estimate the formation heights of these CLASP2/2.1 additional spectral lines and to demonstrate 
their suitability to determine the magnetic field stratification from the photosphere to the upper chromosphere.

%We apply the Weak Field Approximation (WFA) to all the spectral lines 
%in the spectral window between 279.30 and 280.68~nm 
%to infer the longitudinal component of the magnetic field 
%at different layers of the solar atmosphere, from the photosphere to the base of the corona.
%This near-UV spectral region is ideal for studying the magnetism 
%in the solar chromosphere, as demonstrated by the results 
%obtained by the Chromospheric LAyer SpectroPolarimeter missions, CLASP2 and CLASP2.1.
%The spectropolarimetric observations in the \ion{Mg}{2} h and k 
%resonant doublet, \ion{Mn}{1} 279.91 and 280.19~nm resonant lines, 
%and \ion{Fe}{2} 279.79 and 280.66~nm lines obtained 
%by these suborbital space experiments have demonstrated 
%their potential for magnetic field stratification inference in the solar chromosphere. 
%However, several lines of the CLASP2/2.1 spectral region 
%with significant circular polarization signals had remained unexplored.
%After identifying two \ion{Ni}{1} (279.95 and 280.59~nm), one \ion{Mn}{2} (280.62~nm), 
%and one \ion{Fe}{1} (280.53~nm) lines, here we apply the WFA to the spectropolarimetric 
%observations of active region plages by CLASP2 and CLASP2.1.
%By comparing the results with previous studies, 
%we are able to estimate the formation heights of the CLASP2/2.1 lines and to demonstrate 
%their suitability to determine the magnetic field stratification from the photosphere to the upper chromosphere.

\end{abstract}


%%%%%%%%%%%%%%%%%%%%%%%%%%%%%%%%%%%%%%%%%%%%%%%%%%%%%%%%%%%%%%%%%%%%%%%%%%%%%%%%%%%%%%%%%%%%%%%%%
%%%%%%%%%%%%%%%%%%%%%%%%%%%%%%%%%%%%%%%%%%%%%%%%%%%%%%%%%%%%%%%%%%%%%%%%%%%%%%%%%%%%%%%%%%%%%%%%%

\section{Introduction} \label{sec:intro}

Ultraviolet (UV) spectropolarimetry has proved to be very suitable for studying the magnetic field
in the solar chromosphere \citep[see the review by][]{TrujilloBueno2022}.
Spectral lines carry information about the physical conditions 
of the plasma in the region of the stellar atmosphere where they originate and, 
in particular, their polarization is sensitive to the magnetic field strength and orientation. 
This makes the polarization of spectral lines a key tool to infer information 
about the magnetic field in the solar atmosphere.
In the near UV range of the solar spectrum, the strongest spectral lines are the \ion{Mg}{2} h and k 
resonant doublet at 280.35 and 279.64~nm, respectively, 
which are sensitive to the magnetic field near the base of the solar corona.

The promising results of several theoretical studies on the formation and magnetic sensitivity of the \ion{Mg}{2}
h and k lines \citep[][]{Belluzzi-TrujilloBueno2012,AlsinaBallester2016,DelPinoAleman2016,delPinoAleman2020}
motivated the development of the Chromospheric LAyer SpectroPolarimeter 
suborbital mission \citep[CLASP2,][]{Narukage2016,Song2018}.
The unprecedented spectropolarimetric data acquired demonstrated 
the suitability of these spectral lines to map the stratification 
of the magnetic field in the chromosphere of an active region plage.
\cite{Ishikawa2021a} applied the weak-field approximation 
(WFA) to the \ion{Mn}{1} and \ion{Mg}{2} resonant lines 
to infer the longitudinal component of the magnetic field ($B_{\rm L}$) 
in different layers of the solar chromosphere. 
\cite{Afonso2023c} highlighted the suitability of adding two \ion{Fe}{2} 
spectral lines (at 279.79 and 280.66~nm) to this analysis 
to infer information about the magnetic field in deeper layers of the solar atmosphere.
\cite{Li2023} applied the HanleRT-TIC \citep[][]{Li2022,DelPinoAleman2016,delPinoAleman2020} 
to the observed \ion{Mg}{2} h and k profiles, 
inferring the stratification of the thermodynamic parameters of the solar chromospheric plasma and $B_{\rm L}$.
This result allowed to estimate the energy that could be carried by Alfv\'en waves 
propagating in the chromosphere of the observed active region plage and its surrounding enhanced network.

The success of the CLASP2 mission motivated 
a second flight of similar characteristics, dubbed CLASP2.1. 
This suborbital space experiment performed a slit-scan of an active region, 
obtaining 2D spectropolarimetric data in the same spectral window as the previous CLASP2 mission.
\cite{Li2024} applied the HanleRT-TIC to the CLASP2.1 \ion{Mg}{2}
h and k profiles, inferring the stratification of the temperature, electron density,
line-of-sight (LOS) velocity, and $B_{\rm L}$ at each point in the observed field of view.
Additionally, applying the WFA as in \cite{Ishikawa2021a}, 
it was possible to estimate the longitudinal component of the magnetic field in 
the observed active region at different heights in the plage atmosphere (Ishikawa et al. 2025; submitted).

In addition to the \ion{Mg}{2} and \ion{Mn}{1} spectral lines, 
there are a number of spectral lines with significant circular 
polarization signals in the CLASP2 and CLASP2.1 data which remain unexplored.
%In this work, we identify four additional spectral lines, \ion{Ni}{1} 279.95 and 280.59~nm, 
%\ion{Fe}{1} 280.53~nm, and \ion{Mn}{2} 280.62~nm, and study their suitability 
%to infer the magnetic field throughout the solar atmosphere. 
%We apply the WFA to each of these lines and compare the results with the $B_{\rm L}$
%inferred from the \ion{Mg}{2} h and k, the \ion{Mn}{1} 279.91/280.19~nm, and the \ion{Fe}{1} 630.15/630.25~nm lines \citep{Ishikawa2021a}, as well as from the \ion{Fe}{2} 279.79/280.66~nm lines \citep{Afonso2023c}.
In this work, we identify four additional spectral lines, \ion{Ni}{1} 279.95 and 280.59~nm, 
\ion{Fe}{1} 280.53~nm, and \ion{Mn}{2} 280.62~nm, and apply the WFA to each of them. We then compare the results with the $B_{\rm L}$
inferred from the \ion{Mg}{2} h and k, the \ion{Mn}{1} 279.91/280.19~nm, and the \ion{Fe}{1} 630.15/630.25~nm lines \citep{Ishikawa2021a}, as well as from the \ion{Fe}{2} 279.79/280.66~nm lines \citep{Afonso2023c}.
The comparison allows us to make an educated guess about the height of formation of such additional spectral lines and to discuss their suitability to obtain complementary information about the magnetism of the lower layers 
of the solar chromosphere.
We leave for future works the multi-level radiative transfer calculations needed for a detailed study of the formation of these additional spectral lines in solar atmospheric models.

In this work we investigate the suitability of all the above-mentioned spectral lines in the CLASP2/2.1 spectral window (279.30~-~280.68~nm) to infer the stratification of the magnetic field in the solar atmosphere.
In Sec.~\ref{sec:observations} and Sec.~\ref{sec:WFA} we briefly describe the observations and approximations used in this work.
In Sec.~\ref{sec:LineIdentification} we present the identification of all the unexplored spectral lines with significant circular polarization signals.
In Sec.~\ref{sec:CLASP2WFA} and Sec.~\ref{sec:CLASP2.1WFA} we analyze the $B_{\rm L}$ obtained by applying the WFA to the active region plage observed by CLASP2 and CLASP2.1, respectively.  
We summarize our conclusions in Sec.~\ref{sec:conclusions}.

%%%%%%%%%%%%%%%%%%%%%%%%%%%%%%%%%%%%%%%%%%%%%%%%%%%%%%%%%%%%%%%%%%%%%%%%%%%%%%%%%%%%%%%%%%%%%%%%%
%%%%%%%%%%%%%%%%%%%%%%%%%%%%%%%%%%%%%%%%%%%%%%%%%%%%%%%%%%%%%%%%%%%%%%%%%%%%%%%%%%%%%%%%%%%%%%%%%

\section{Observations} \label{sec:observations}
The spectropolarimetric data used in this work were acquired by the CLASP2 and CLASP2.1 suborbital missions 
on 11 April 2019 and 8 October 2021, respectively.

The CLASP2 mission observed  with a 196~arcsec spectrograph slit in two fixed positions: a quiet Sun region close to the solar limb, 
and the east side of the active region NOAA~12738.
The active region observation was performed between 16:53:40 and 16:56:16~UT, i.e., 
a total exposure time of 156~s \citep{Ishikawa2021a}. 
In this work, we use the temporally averaged signals,
which achieve a polarimetric sensitivity at the continuum of Stokes $V$ of $2\cdot10^{-4}$ relative to the intensity at the line center of \ion{Mg}{2} k  \citep{Song2022}. 

The CLASP2.1 mission performed a slit-scan of the active region NOAA 12882.
The observation was carried out between 17:42:13 and 17:47:38~UT. 
During this time, the instrument acquired data at 16 slit positions,
with an exposure time for each position of about 18~s. 
The slit raster step is 1.8~arcsec on average. 
Due to the reduced exposure time of each slit scan, the polarimetric sensitivity at the continuum of Stokes V is about $10^{-3}$ relative to the line center of  \ion{Mg}{2} k, worse than in the case of the CLASP2 observations.

Both missions covered the spectral window between ~279.30 and ~280.68~nm, 
with a spectral sampling of 49.8~m\AA/pixel,
and a spectral point spread function which can be approximated by a Gaussian with a full width at half maximum of about 110~m\AA\ \citep{Song2018, Yoshida2018, Tsuzuki2020}. 
The spatial sampling along the slit is 0.53~arcsec/pixel. 

\section{Weak-field approximation} \label{sec:WFA}
The WFA provides a very fast method to infer information about the magnetic field in the solar atmosphere, in contrast to the usually slow non-local thermodynamic (NLTE) radiative transfer inversions needed for chromospheric spectral lines. 

The WFA  provides an analytical solution to the radiative transfer equations when some assumptions are satisfied \citep[][]{LL04}. These assumptions are that $B_{\rm L}$ is constant in the formation region of the line and that the magnetic field strength must be weak enough to guarantee that the Zeeman splitting ($\Delta\lambda_B = 4.6686\cdot10^{-13}B\lambda_0^2$, with $B$ the magnetic field strength in gauss and $\lambda_0$ the line center's wavelength in \AA) is significantly smaller than the Doppler width ($\lambda_D = \frac{\lambda_{0}}{c} \sqrt{\frac{2kT}{m} + v_m^2}$, with $T$ and $v_m$
the temperature and turbulent velocity in the formation region of the line, respectively, $c$ the speed of light, 
$k$ the Boltzmann constant, and $m$ the mass of the atom) of the line; i.e.:
\begin{equation}
g_{eff}\frac{\Delta\lambda_B}{\Delta\lambda_D} << 1,
\end{equation}

When these assumptions are satisfied, we can approximate the circular polarization Stokes $V$ profile as
\begin{equation}\label{eq:wfa}
    V(\lambda) = -4.6686\cdot10^{-13}g_{eff}\lambda_0^2B_{\rm L}\frac{\partial I(\lambda)}{\partial \lambda}.
\end{equation}

%%%%%%%%%%%%%%%%%%%%%%%%%%%%%%%%%%%%%%%%%%%%%%%%%%%%%%%%%%%%%%%%%%%%%%%%%%%%%%%%%%%%%%%%%%%%%%%%%
%%%%%%%%%%%%%%%%%%%%%%%%%%%%%%%%%%%%%%%%%%%%%%%%%%%%%%%%%%%%%%%%%%%%%%%%%%%%%%%%%%%%%%%%%%%%%%%%%
% 

\section{Spectral line identification} \label{sec:LineIdentification}

\begin{figure*}[ht!]
    \centering
    \includegraphics[width=1.\textwidth]{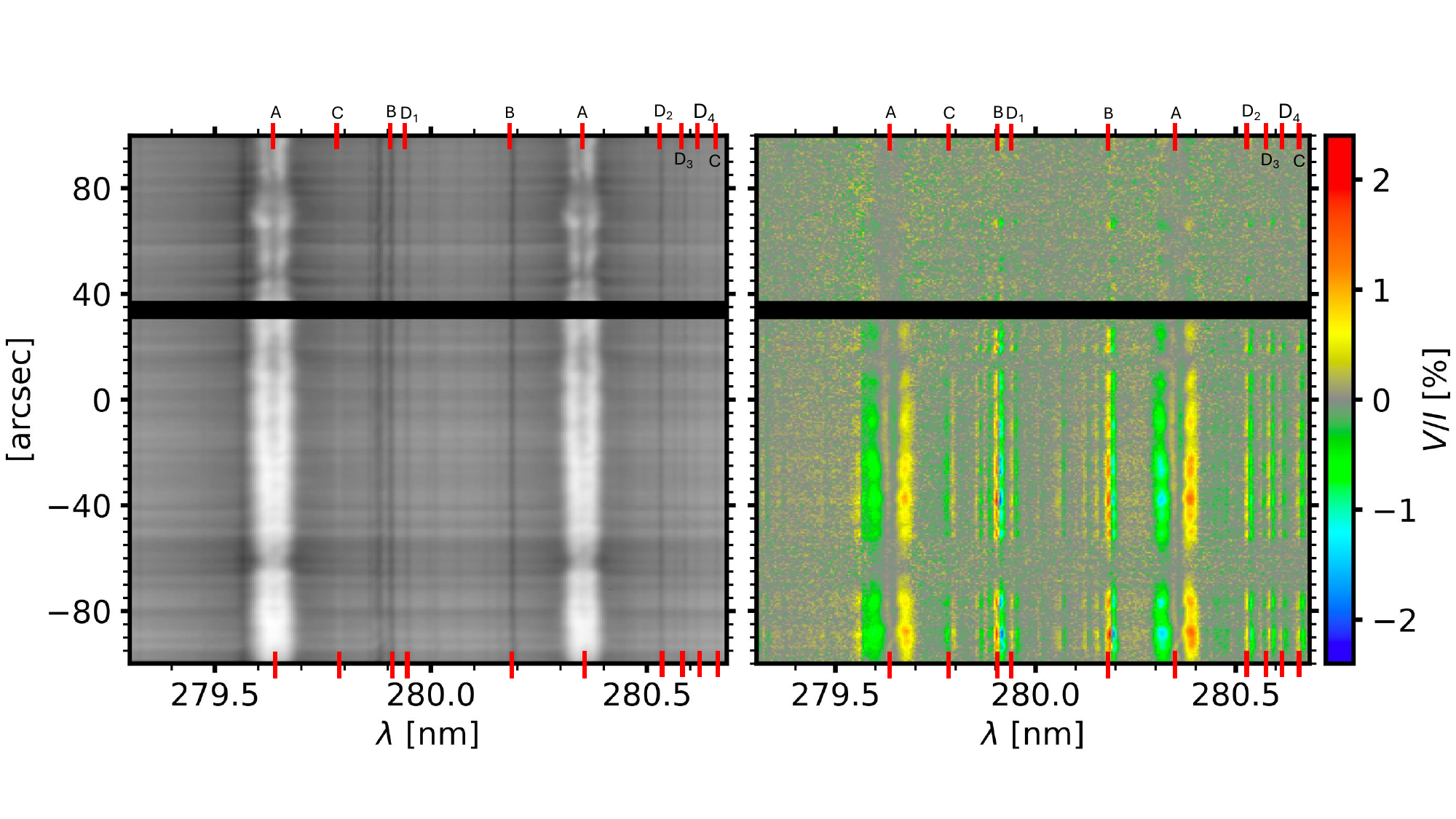}
    \caption{Intensity (left) and fractional circular polarization (right) spectra along the CLASP2 slit. 
    Red lines on the wavelength axis indicate the positions of all the spectral lines with significant circular polarization signals. 
    The A and B labels indicate the \ion{Mg}{2} h and k lines and the  \ion{Mn}{1} resonant lines (at 279.91 and 280.19~nm) analyzed in \cite{Ishikawa2021a}, respectively. The C label indicates the position of the \ion{Fe}{2} lines (at 279.79 and 280.66~nm) analyzed in \cite{Afonso2023c}.
    The D$_{1-4}$ labels indicate the wavelength of the 4 lines analyzed in this work, respectively: \ion{Ni}{1} 279.95~nm, \ion{Fe}{1} 280.53~nm, \ion{Ni}{1} 280.59~nm, and \ion{Mn}{2} 280.62~nm.
    The wavelengths are vacuum wavelengths.
    }
    \label{fig:lines}
\end{figure*}

In addition to the spectral lines analyzed in \cite{Ishikawa2021a} and \cite{Afonso2023c}, 
in the CLASP2/CLASP2.1 spectral region we find four additional spectral lines with circular polarization signals larger than 0.2~\%: 
\ion{Ni}{1} 279.95 and 280.59~nm, \ion{Fe}{1} 280.53~nm, and \ion{Mn}{2} 280.62~nm.
Figure~\ref{fig:lines} shows all the atomic lines showing substantial circular polarization signals in the CLASP2/2.1 spectral window.
 In this and the remaining figures we use vacuum wavelengths.
These spectral lines generally show weak absorption profiles in intensity in the CLASP2 and CLASP2.1 observations.
The absorption intensity profile of the strongest of these lines (i.e., \ion{Fe}{1} 280.53~nm) is similar in depth to the \ion{Mn}{1} resonant lines. 
The strongest Stokes $V/I$ signal is of about 0.5$\%$, which corresponds to the \ion{Fe}{1} 280.53~nm line. 

In Table~\ref{tab:newLines} we show the atomic data of these spectral lines, 
including their effective Land\'e factor (showing values between 1.08 and 1.61) computed by assuming LS coupling and using the experimental Land\'e factor of each individual atomic level. 
The atomic data and experimental Land\'e factors have been taken from the National Institute of Standards and Technology (NIST) Atomic Spectra database \citep{NIST_ASD}.

\begin{deluxetable*}{cccccc} 
\tablecaption{ Ion, vacuum wavelength, atomic configuration, energies, Einstein coefficient of spontaneous emission, and effective Landé factor for the additional atomic transitions identified in the CLASP2/2.1 spectral window.}

\tablehead{
\colhead{Ion} & \colhead{$\lambda_{vacuum}$~[nm]} & \colhead{Transition} & \colhead{Levels energy~[cm$^-1$]} & \colhead{A$_{ul}$~[s$^{-1}$]} & \colhead{g$_{eff}$} }

\startdata
\ion{Ni}{1}  & 279.95 & $^3$D$_{2}$ - $^1$D$_{2}^{\circ}$  & \,\, 880 - 36601 & $5.8\cdot 10^6$ & 1.08 \\
\ion{Fe}{1}  & 280.53 & $^5$F$_{4}$ - $^5$G$_{4}^{\circ}$ & \,7377 - 43023 & $1.0\cdot 10^7$ & 1.19 \\
\ion{Ni}{1}  & 280.59 & $^3$F$_{4}$ - $^1$F$_{3}^{\circ}$  & \:\:\:\:\:\:\:\:0 - 35639 & $8.8\cdot 10^5$ & 1.61 \\
\ion{Mn}{2} & 280.62 & $^3$G$_{4}$ - $^3$G$_{4}^{\circ}$  &34911 - 70546 & $7.1\cdot 10^7$ & 1.11 \\ 
\enddata
\end{deluxetable*} \label{tab:newLines}

\begin{figure*}[ht!]
    \centering
    \includegraphics[width=0.8\textwidth]{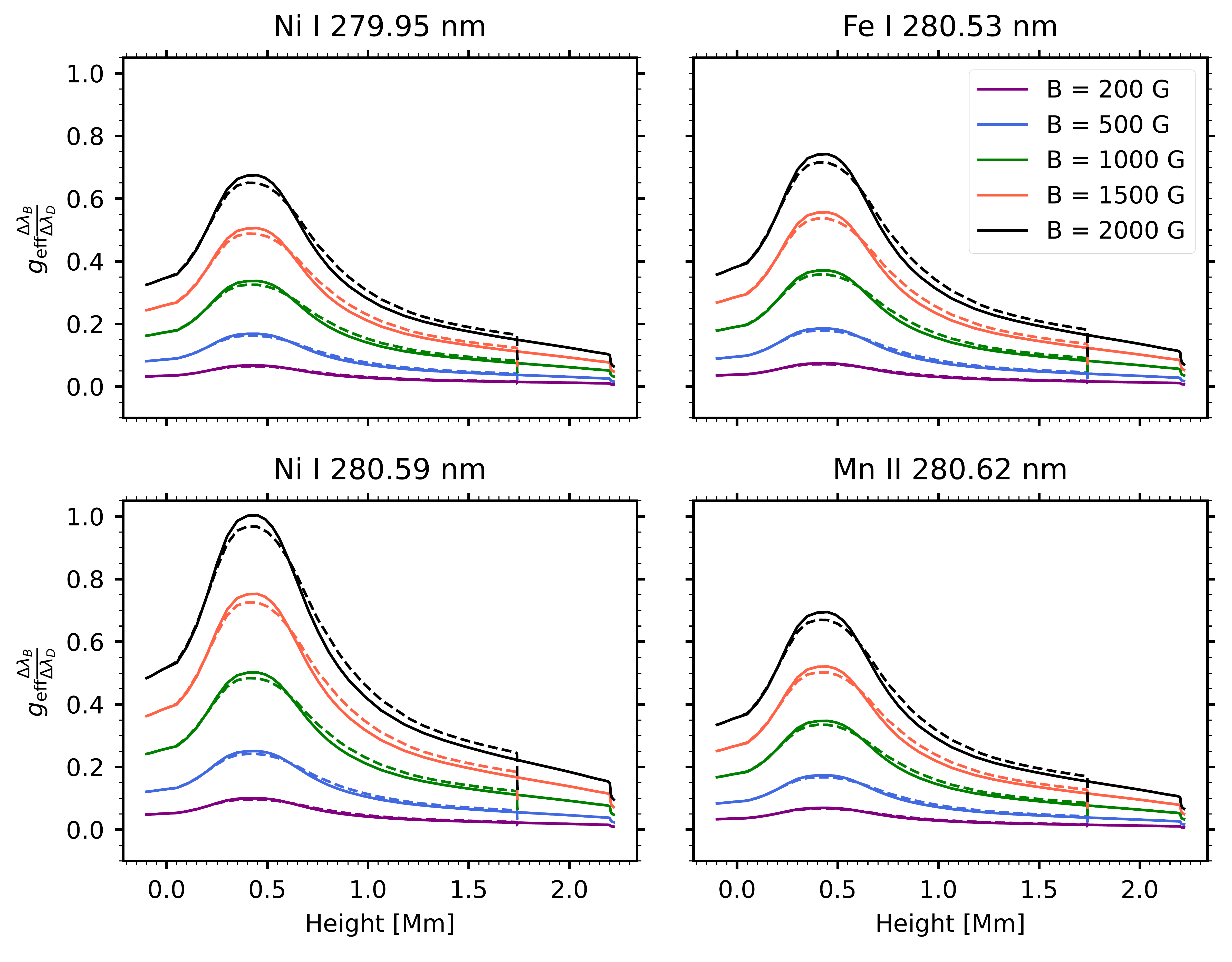}
    \caption{Variation with height of the ratio $R=g_{\rm eff}\Delta\lambda_B/\Delta\lambda_D$ in the FAL-C (solid curves) and FAL-P (dashed curves) atmospheric models, assuming uniform magnetic fields with the strengths indicated in the upper-right panel. The different panels correspond to the spectral lines listed in Table~\ref{tab:newLines}, indicated at the top of each panel.}
    \label{fig:6Zeeman}
\end{figure*}

Figure~\ref{fig:6Zeeman} shows the variation with height of the ratio $g_{\rm eff}\Delta\lambda_B/\Delta\lambda_D$
in the C (representative of the quiet Sun, hereafter FAL-C) and P (representative of a plage region, hereafter FAL-P) models of \cite{Fontenla1993}, 
for a uniform magnetic field with strengths of 200, 500, 1000, and 2000~G.
As mentioned in Sec.~\ref{sec:WFA}, the suitability of the WFA requires that $R=g_{\rm eff}\Delta\lambda_B/\Delta\lambda_D\ll1$. 
However, based on numerical testing in plane-parallel models of the solar atmosphere, in practice $R = 0.5$ tends to be a suitable upper limit for many solar applications, including the present study.
By assuming the worst-case scenario of the spectral line forming at the temperature minimum ($\approx0.5$~Mm), the WFA would be suitable for magnetic field strengths below 1500~G.
For the \ion{Ni}{1} 280.59~nm line, the condition is more restrictive and the WFA would be suitable for magnetic field strengths below 1000~G. 

For the above-mentioned four lines, the WFA thus seems suitable in most regions of the solar atmosphere 
%. We do not expect the WFA to be applicable
except in sunspots, where we expect larger magnetic field strengths on the order of kilogauss.

%%%%%%%%%%%%%%%%%%%%%%%%%%%%%%%%%%%%%%%%%%%%%%%%%%%%%%%%%%%%%%%%%%%%%%%%%%%%%%%%%%%%%%%%%%%%%%%%%
%%%%%%%%%%%%%%%%%%%%%%%%%%%%%%%%%%%%%%%%%%%%%%%%%%%%%%%%%%%%%%%%%%%%%%%%%%%%%%%%%%%%%%%%%%%%%%%%%

\section{Application of the WFA to the CLASP2 data} \label{sec:CLASP2WFA}

Previous analysis of the active region plage observed by CLASP2 proved that the application of the WFA to the \ion{Mg}{2} and \ion{Mn}{1} resonance lines allows the determination of $B_{\rm L}$ at different layers of the solar chromosphere. From the inner lobes of the Stokes $V$ profiles of the \ion{Mg}{2} h and k spectral lines, we can determine $B_{\rm L}$ in the upper layers of the chromosphere, just below the transition region; from the outer lobes of \ion{Mg}{2} h, we can infer $B_{\rm L}$ in the middle chromosphere, and with the resonant \ion{Mn}{1} spectral lines we can infer $B_{\rm L}$ in the lower chromosphere \cite[see][]{Ishikawa2021a,delPinoAleman2022}. Likewise, it has been shown that the \ion{Fe}{2} 280.66~nm line is suitable for inferring $B_{\rm L}$ in the upper photosphere, while the height of formation of the \ion{Fe}{2} line at 279.79~nm varies from the upper photosphere in the quiet Sun to the middle chromosphere in plage regions \citep{Afonso2023c}.

\begin{figure*}[htp!]
    \centering
    \includegraphics[width=1.\textwidth]{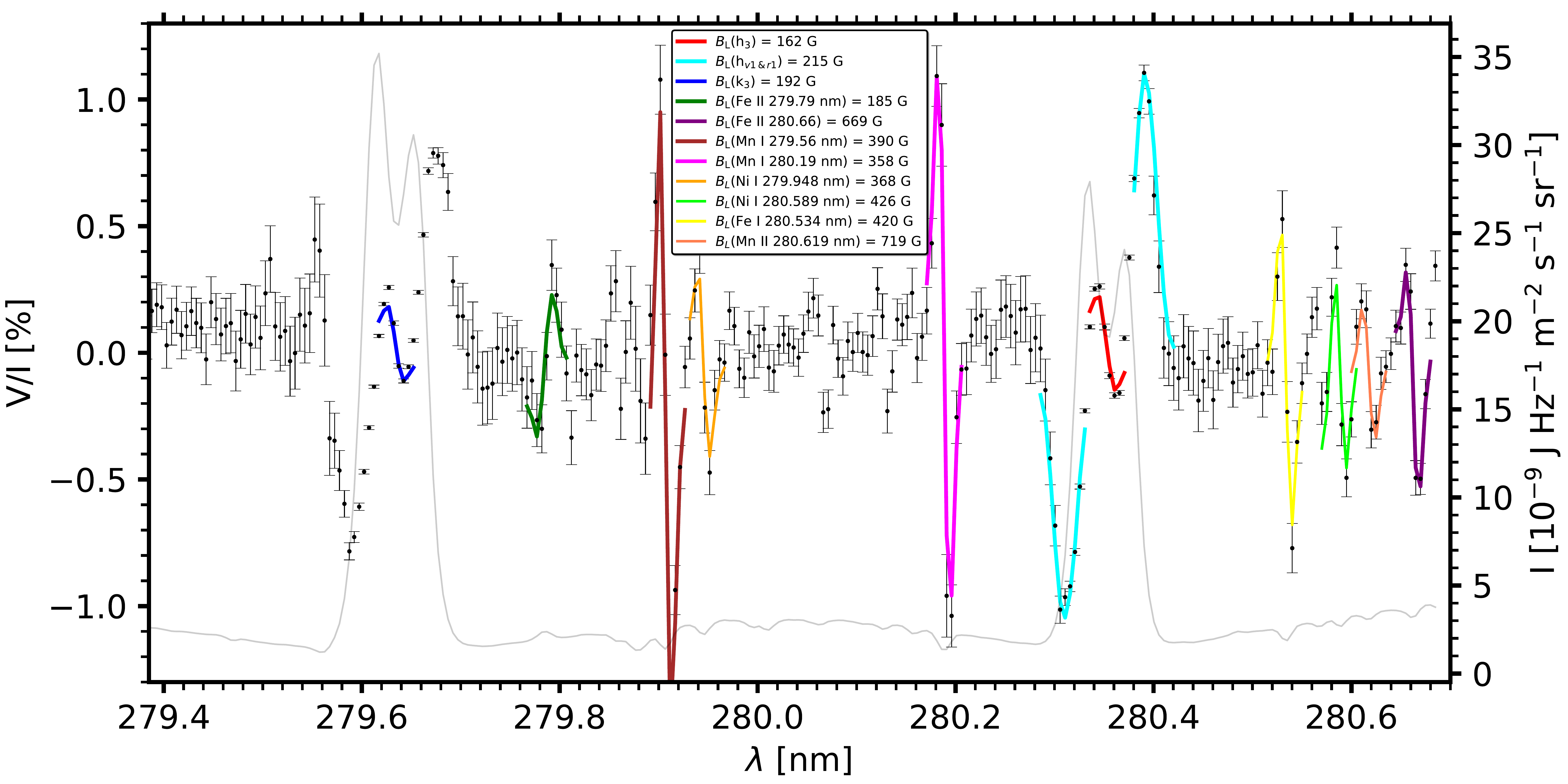}
    \caption{The fractional circular polarization (black dots, with error bars) of the CLASP2 plage target at the pixel corresponding to the location at -24.12~arcsec in Fig.~\ref{fig:lines}. 
    The wavelengths are vacuum wavelengths.
    The thin gray curve is the corresponding intensity profile. Color curves show the WFA fits for each of the spectral lines analyzed in this work. The legend in the inset indicates both the spectral line and the resulting $B_{\rm L}$ value.}
    \label{fig:CLASP2wfaProf}
\end{figure*}

In this section we apply the WFA to all the spectral lines of the CLASP2 spectral region with significant circular polarization signals, including  
those highlighted in Table~\ref{tab:newLines}, to infer $B_{\rm L}$ in different layers of the solar atmosphere. In Fig.~\ref{fig:CLASP2wfaProf} we show the inferred $B_{\rm L}$ and the corresponding WFA fit in a given pixel.
The selected pixel is located in the brightest area of the observed active region plage (-24.12~arcsec in Fig.~\ref{fig:lines}). The fractional circular polarization signals of the 
four spectral lines indicated in Table~\ref{tab:newLines}
are about or below 0.5~\%. Although the signal-to-noise ratio in these lines is relatively low, we obtain reasonably good fits when applying the WFA.

The $B_{\rm L}$ values inferred from the \ion{Mg}{2}, \ion{Mn}{1}, and \ion{Fe}{2} spectral lines suggest that the magnetic field strength decreases with height, in agreement with the results from \cite{Ishikawa2021a} and \cite{Afonso2023c}. The weakest $B_{\rm L}$ values were inferred from the inner lobes of \ion{Mg}{2} h and k, associated to the upper chromosphere. It is important to note that the instrumental degradation produced by the CLASP2 finite spectral resolution systematically overestimates the values inferred through the application of the WFA to the inner lobes of \ion{Mg}{2} h and k, but this error is not significant for the inferred values in the CLASP2 data (see Appendix~\ref{sec:appA} for further details). 
The largest $B_{\rm L}$ values were inferred from the \ion{Fe}{2} 280.66~nm photospheric line. 
The $B_{\rm L}$ inferred from the \ion{Mn}{1} resonant lines, from the \ion{Mg}{2} h outer lobes, and from the \ion{Fe}{2} 279.79~nm line are in between these extremes, as expected because these lines originate in the lower and middle chromosphere.

\begin{figure*}[htp!]
    \centering
    \includegraphics[width=1\textwidth]{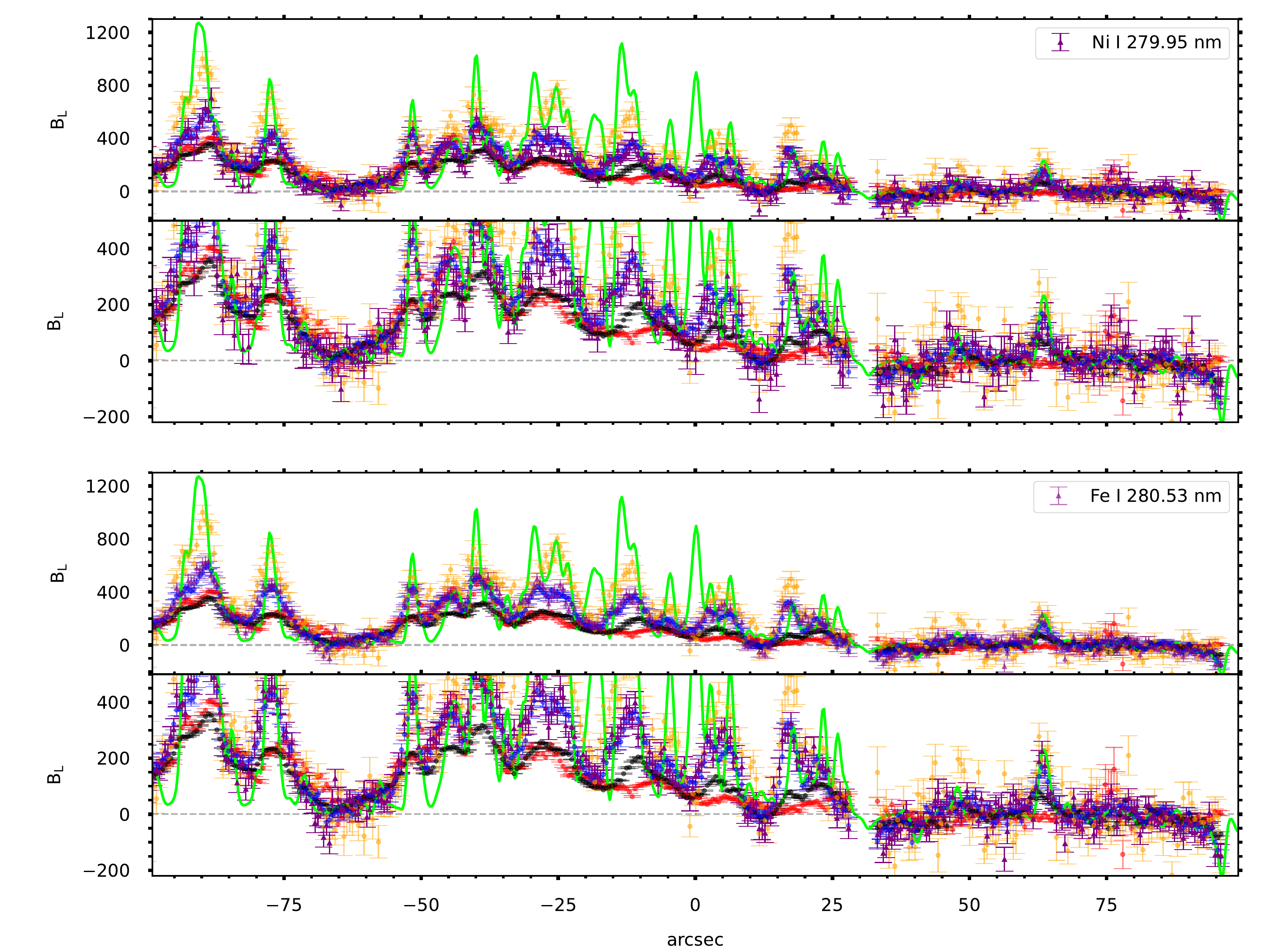} 
    \caption{Inferred $B_{\rm L}$ at each position along the CLASP2 spectrograph slit. The colored circles represent the $B_{\rm L}$ values inferred by the application of the WFA to different spectral regions: red circles for the average value inferred from the inner lobes of \ion{Mg}{2} h and k, black circles for the outer lobes of \ion{Mg}{2} h, blue circles for the \ion{Mn}{1} resonant lines, and orange circles for the \ion{Fe}{2} 280.66~nm line. The purple triangles show the $B_{\rm L}$ inferred from the \ion{Ni}{1} 279.95~nm line (top panel) and from the \ion{Fe}{1} 280.53~nm line (bottom panel). The light-green curve represents the values obtained from the Milne-Eddington inversion of the Hinode/SOT-SP observation of the \ion{Fe}{1} 630.15/630.25~nm spectral lines \citep{Ishikawa2021a}.  
    The lower half of each panel shows a zoom of the $B_{\rm L}$ axis.}
    \label{fig:CLASP2wfa}
\end{figure*}

\begin{figure*}[htp!]
    \centering
    \includegraphics[width=1\textwidth]{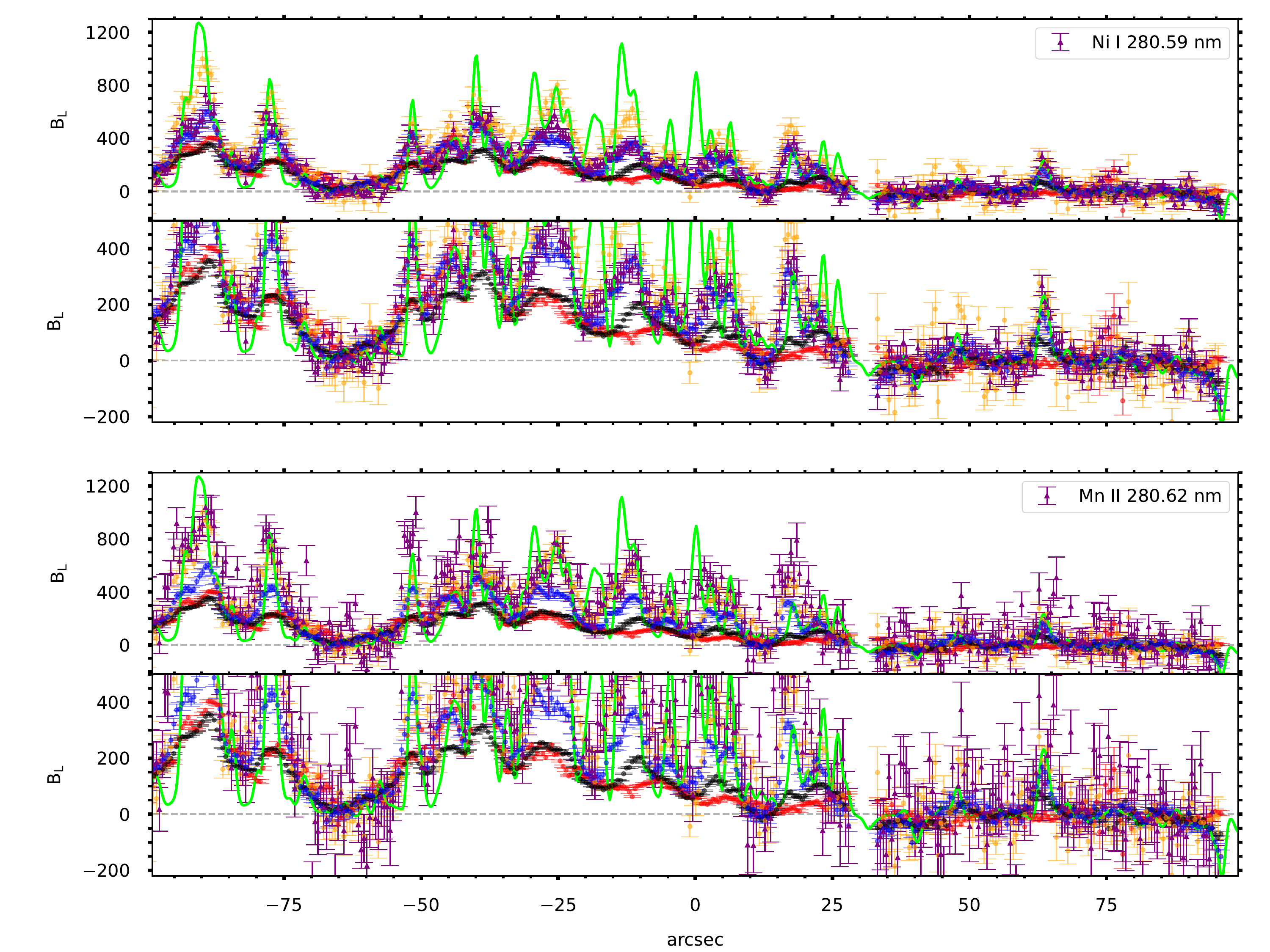} 
    \caption{Same as Fig.~\ref{fig:CLASP2wfa}, but for the \ion{Ni}{1} 280.59~nm line (top panel) and the \ion{Mn}{2} 280.62~nm line (bottom panel).}
    \label{fig:CLASP2wfa2}
\end{figure*}

Regarding the spectral lines identified in this work, the $B_{\rm L}$ inferred from the \ion{Ni}{1} 279.95 and 280.59~nm lines, and from the \ion{Fe}{1} 280.53~nm line (between 358 and 426~G) are very close to the values inferred from the \ion{Mn}{1} resonant lines (358 and 390~G). The $B_{\rm L}$ inferred from the \ion{Mn}{2} 280.62~nm line is the largest (719~G) among all the  lines in the CLASP2 window. Although we do not have information on their formation heights (e.g., from radiative transfer calculations), we can make reasonable assumptions by comparing the inferred $B_{\rm L}$ values with those obtained from the \ion{Mg}{2} and \ion{Mn}{1} lines.

Figures~\ref{fig:CLASP2wfa} and \ref{fig:CLASP2wfa2} show the variation of the inferred $B_{\rm L}$ along the spatial direction of the CLASP2 slit in the plage target. 
The errors in $B_{\rm L}$ were computed following the methodology described in \cite{MartinezGonzalez2012}.
The inferred values are close and compatible with zero in the quiet Sun regions (between 35 and 95~arcsec in Figs.~\ref{fig:lines},~\ref{fig:CLASP2wfa}, and \ref{fig:CLASP2wfa2}, except for the crossing with the network at about 65~arcsec), considering the uncertainty due to the low signal-to-noise ratio in these spectral lines observvations. In the slit positions corresponding to the active region plage ($\lesssim25$~arcsec), the $B_{\rm L}$ inferred from the four additional spectral lines follows a pattern similar to that of the $B_{\rm L}$ corresponding to the \ion{Mg}{2}, \ion{Mn}{1}, and \ion{Fe}{2} lines. These results support the suitability of applying the WFA to these spectral lines to infer  $B_{\rm L}$.

The $B_{\rm L}$ inferred from the \ion{Ni}{1} 279.95 and  280.59~nm lines, and from the \ion{Fe}{1} 280.53~nm line, are close to the $B_{\rm L}$ inferred from the \ion{Mn}{1} resonant lines. Consequently, we estimate that the formation region of these lines is similar to that of the \ion{Mn}{1} lines, i.e., they form somewhere in the lower chromosphere. 

At all slit positions in the active region plage, the inferred $B_{\rm L}$ from the \ion{Mn}{2} 280.62~nm is notably larger than for any other line, including the \ion{Fe}{2} 280.66~nm line, which originates in the upper layers of the photosphere \citep[][]{Afonso2023c}.
It is important to note that the signal-to-noise ratio in this spectral line is the lowest and, consequently, the uncertainty in the inferred $B_{\rm L}$ is considerably larger than for any of the other considered lines. 
Nevertheless, this result is reliable enough to conclude that the \ion{Mn}{2} 280.62~nm spectral line forms in the solar photosphere, below the \ion{Fe}{2} 280.66~nm line and at heights similar to those of the \ion{Fe}{1} 630.15/630.25~nm lines observed by Hinode/SOT-SP.

%%%%%%%%%%%%%%%%%%%%%%%%%%%%%%%%%%%%%%%%%%%%%%%%%%%%%%%%%%%%%%%%%%%%%%%%%%%%%%%%%%%%%%%%%%%%%%%%%
%%%%%%%%%%%%%%%%%%%%%%%%%%%%%%%%%%%%%%%%%%%%%%%%%%%%%%%%%%%%%%%%%%%%%%%%%%%%%%%%%%%%%%%%%%%%%%%%%
\section{Application of the WFA to the CLASP2.1 data} \label{sec:CLASP2.1WFA}

\begin{figure*}[htp!]
    \centering
    \includegraphics[width=1.\textwidth]{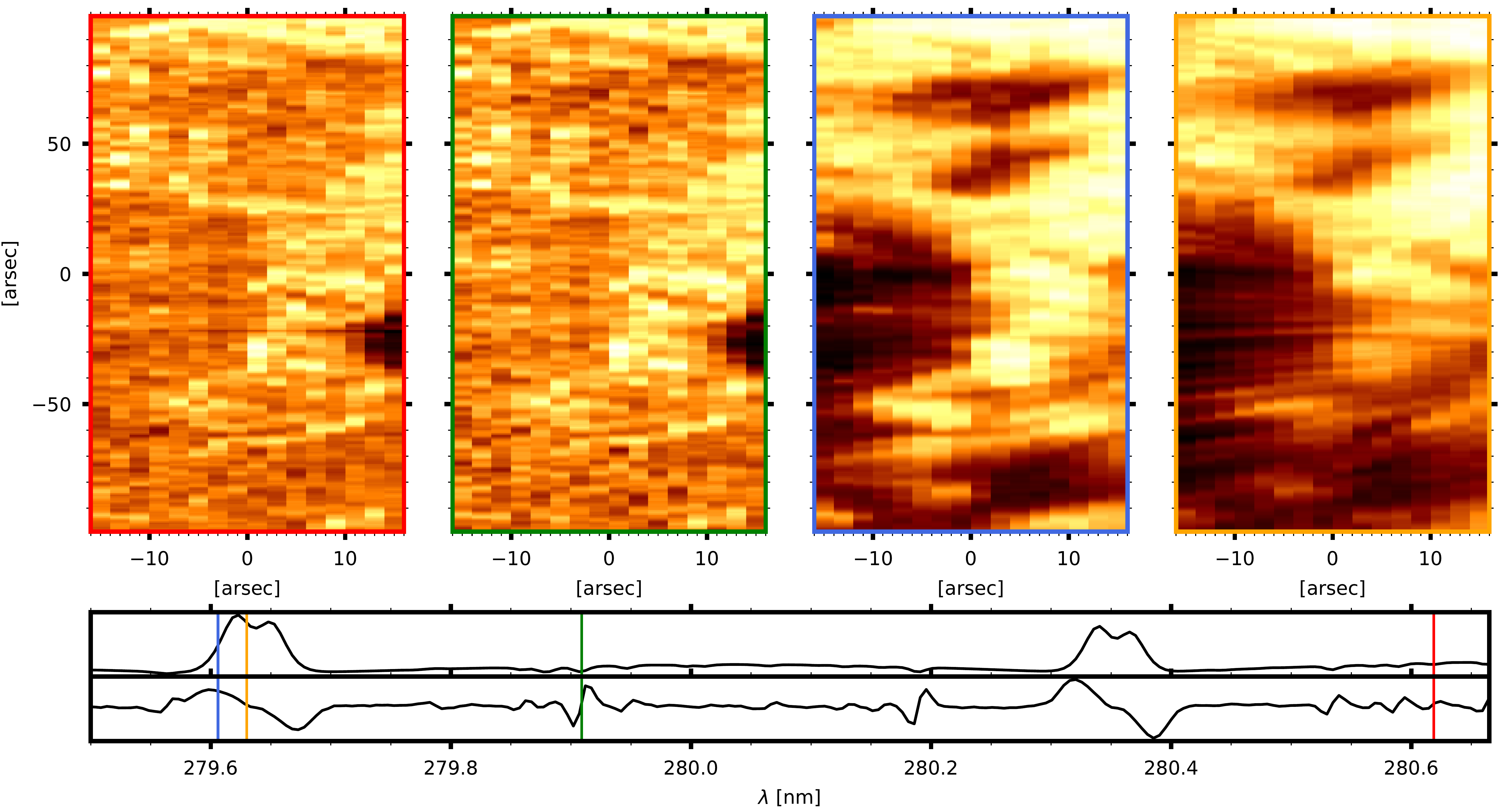}    
    \caption{
     Intensity maps (upper row) at four different wavelengths in the CLASP2.1 observation. The bottom panel shows the spatially averaged intensity (top half) and fractional circular polarization (bottom half) profiles.
     The wavelengths are vacuum wavelengths.
     The color of the axes in the upper row identifies the corresponding wavelength, indicated with a vertical line in the bottom panel, namely red for \ion{Mn}{2} 280.62~nm, green for \ion{Mn}{1} 279.91~nm, blue for the \ion{Mg}{2} k blue outer lobe, and orange for the \ion{Mg}{2} k line core. The panels in the top row are ordered from left to right following the estimated formation height, from lowest to highest.
    }
    \label{fig:CLASP2.1Intensity}
\end{figure*}

The CLASP2.1 mission observed a 2D area of an active region plage by measuring the wavelength variation of the four Stokes paramaters at 16 parallel slit positions with a separation of about 1.8~arcsec. The characteristics of the observations are the same as those from the CLASP2, but with a shorter exposure time for each slit position. The vertical size of the 2D map is 196~arcsec with a spatial resolution of $\sim$0.53~arcsec/pixel. The horizontal size is about 28~arcsec. 

Unfortunately, with the CLASP2.1 data it was not possible to exploit the two \ion{Fe}{2} spectral lines analyzed in \cite{Afonso2023c}.
In the CLASP2.1 observation the Stokes $V$ signal-to-noise ratio of the \ion{Fe}{2} 279.79~nm line is too weak for the application of the WFA.
Moreover, the \ion{Fe}{2} 280.66~nm line was not observed due to the slightly smaller spectral range of CLASP2.1.

The spatial resolution of the CLASP2.1 observations only allows for the identification of few details in the magnetic concentrations of the observed active region. 
However, it is possible to appreciate the expansion of these magnetic concentrations with height. 
In Fig.~\ref{fig:CLASP2.1Intensity} we show four intensity maps corresponding to four different wavelengths in the CLASP2.1 window, with different formation heights. 
The bright structures, usually associated  with stronger magnetic fields, are more widespread and diffuse in the higher layers of the atmosphere, i.e., in the line core (wing) of the \ion{Mg}{2} k line forming in the upper (middle) chromosphere.

The lines originating deeper in the atmosphere, like the \ion{Mn}{1} and \ion{Mn}{2} lines, show more defined bright structures surrounded by a more homogeneous quiet Sun. However, the low spatial resolution does not allow us to see the granulation pattern. We verified that the intensity maps from the \ion{Ni}{1} 279.95 and 280.53~nm lines, and the \ion{Fe}{1} 280.53~nm line are similar to the \ion{Mn}{1} 279.91~nm map, as expected from their similar formation heights.

\begin{figure*}[htp!]
    \centering
    \includegraphics[width=1.\textwidth]{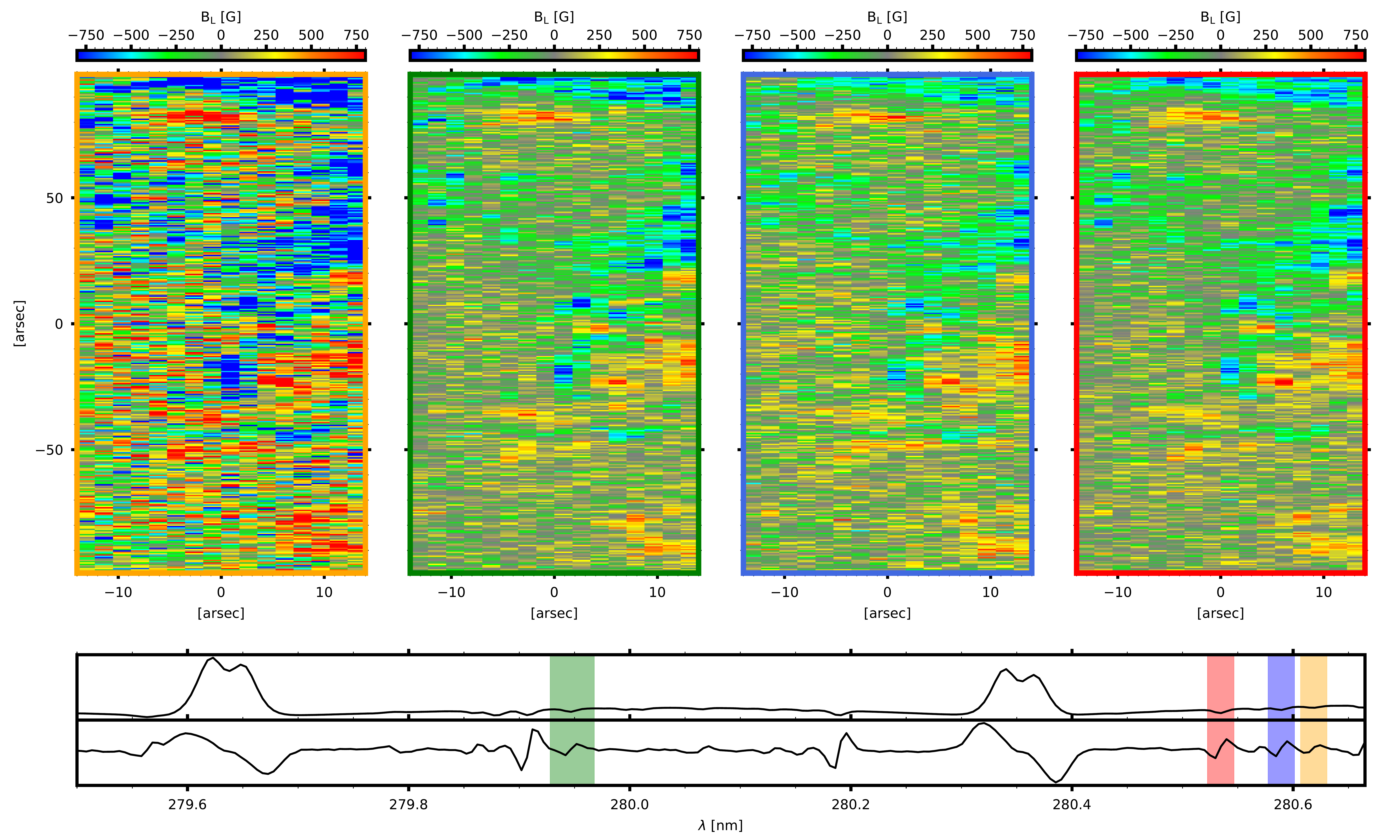}
    \caption{Upper panels: $B_{\rm L}$ maps inferred from the application of the WFA to the spectral lines of  Table~\ref{tab:newLines} in the CLASP2.1 observation. The bottom panel shows the intensity (top half) and the fractional circular polarization (bottom half) profiles.
    The wavelengths are vacuum wavelengths.
    The color of the axes in the upper-row panels identifies the corresponding spectral line for the application of the WFA, indicated with a vertical shaded area in the bottom panel, namely orange for \ion{Mg}{2} 280.62~nm, green for \ion{Ni}{1} 279.95~nm, blue for \ion{Ni}{1} 280.59~nm, and red for \ion{Fe}{1} 280.53~nm. The panels in the top row are ordered from left to right following the estimated formation height, from the lowest to the highest.
    }
    \label{fig:WFACLASP2.1}
\end{figure*}

As in Sec.~\ref{sec:CLASP2WFA}, we apply the WFA to all the spectral lines with significant circular polarization signals. We include in our analysis the $B_{\rm L}$ values inferred from the application of the WFA to the \ion{Mg}{2} and \ion{Mn}{1} spectral lines. 
Figure~\ref{fig:WFACLASP2.1} shows the $B_{\rm L}$ maps obtained from the application of the WFA to the \ion{Mn}{2} 280.62~nm line, the \ion{Ni}{1} 279.95 and 280.59~nm lines, and the \ion{Fe}{1} 280.53~nm line.
Due to the shorter exposure time and ensuing larger uncertainty (compared to the CLASP2 data in Sec.~\ref{sec:CLASP2WFA}), the resulting $B_{\rm L}$ maps are quite noisy in the quiet Sun areas of the field of view. 
We observe large magnetic field concentrations that seem to correlate with the location of the bright regions in the intensity maps of Fig.~\ref{fig:CLASP2.1Intensity}. 
 We can identify mainly four different $B_{\rm L}$ concentrations, four with positive polarity in the lower half of the map (at around -80, -50, -30 and -20~arcsec in the vertical axis) and two with negative polarity in the upper half of the map (at around 30 and 90~arcsec in the vertical axis). 
The strongest $B_{\rm L}$ are inferred from the \ion{Mn}{2} 280.62~nm line, with values around 1300~G in the large magnetic concentration located at around 30~arcsec in the vertical axis.
The maps inferred from the \ion{Ni}{1} and \ion{Fe}{1} lines show similar magnetic field concentrations but with not-so-intense $B_{\rm L}$ values.

\begin{figure*}[htp!]
    \centering
    \includegraphics[width=1.\textwidth]{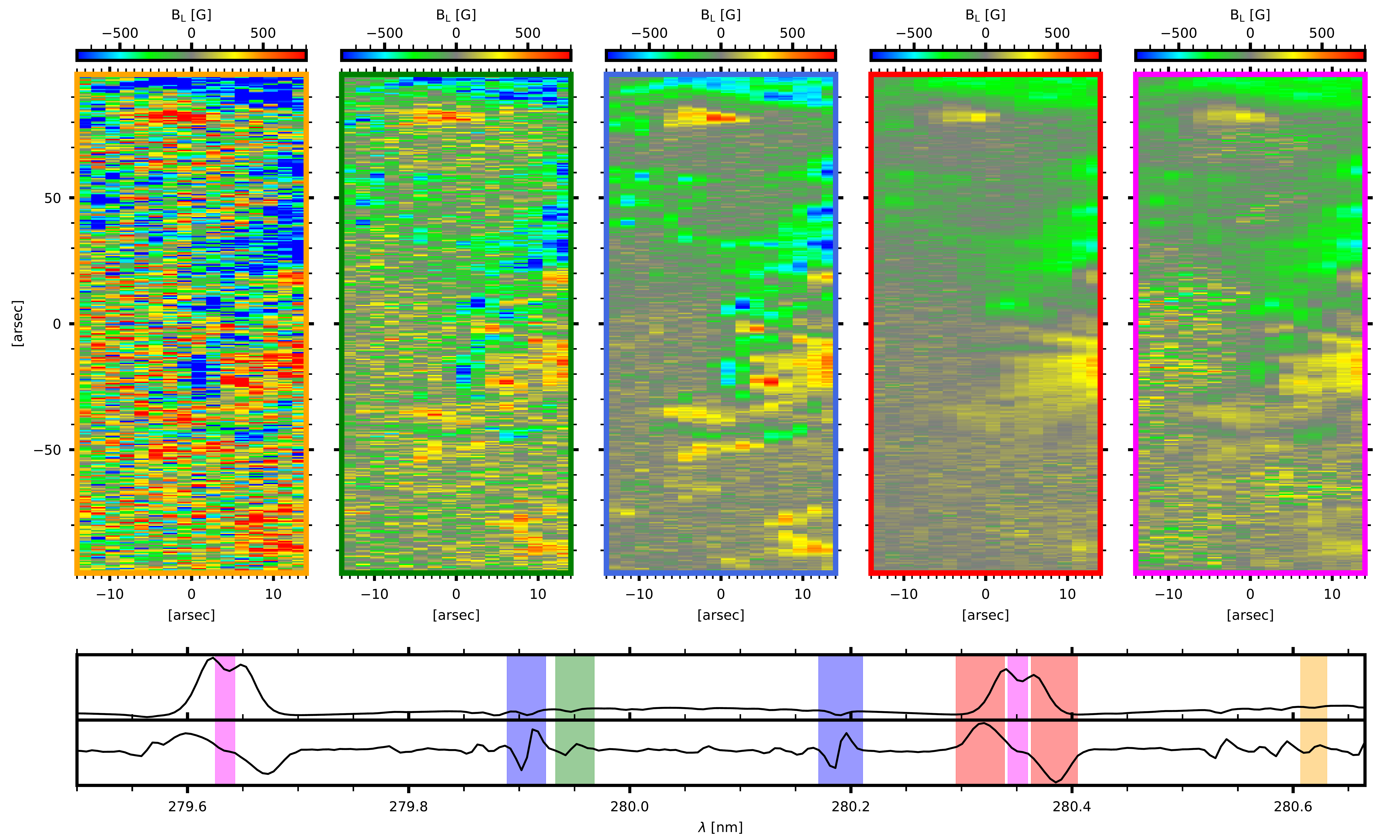}
    \caption{Upper panels: $B_{\rm L}$ maps inferred from the application of the WFA to five spectral regions in the CLASP2.1 observation. The bottom panel shows the spatially averaged intensity (top half) and fractional circular polarization (bottom half) profiles. 
    The wavelengths are vacuum wavelengths.
    The color of the axes in the upper-row panels identifies the corresponding spectral region for the application of the WFA, indicated with a vertical shaded area in the bottom panel, namely orange for \ion{Mn}{2} 280.62~nm, green for \ion{Ni}{1} 279.95~nm, blue for \ion{Mn}{1} 279.91 and 280.19~nm, red for the \ion{Mg}{2} h outer lobes, and magenta for the \ion{Mg}{2} h and k inner lobes. The panels in the top row are ordered from left to right following the estimated formation height, from the lowest to the highest.
    }
    \label{fig:WFACLASP2.1_2}
\end{figure*}

In Fig.~\ref{fig:WFACLASP2.1_2} we show the $B_{\rm L}$ maps inferred from \ion{Mn}{2} 280.62~nm and \ion{Ni}{1} 279.95~nm lines as well as those inferred from the \ion{Mn}{1} resonant lines, the \ion{Mg}{2} h outer lobes, and the \ion{Mg}{2} h and k inner lobes. These results are in agreement with those obtained in Sec.~\ref{sec:CLASP2WFA}. We find the largest $B_{\rm L}$ values in the $B_{\rm L}$ map inferred from the \ion{Mn}{2} 280.62~nm line. 
The $B_{\rm L }$ values inferred from the \ion{Ni}{1} 279.95~nm line show a similar distribution to the one found for the \ion{Mn}{1} lines but with slightly lower values. Despite the noisiness of the measurements, our results are consistent with a $B_{\rm L}$ decreasing with height.

%%%%%%%%%%%%%%%%%%%%%%%%%%%%%%%%%%%%%%%%%%%%%%%%%%%%%%%%%%%%%%%%%%%%%%%%%%%%%%%%%%%%%%%%%%%%%%%%%
 %%%%%%%%%%%%%%%%%%%%%%%%%%%%%%%%%%%%%%%%%%%%%%%%%%%%%%%%%%%%%%%%%%%%%%%%%%%%%%%%%%%%%%%%%%%%%%%%%

\section{Conclusions} \label{sec:conclusions}

In this work we have identified four additional spectral lines in the CLASP2/2.1 spectral region:
\ion{Ni}{1} 279.95 and 280.59~nm, \ion{Fe}{1} 280.53~nm, and \ion{Mn}{2} 280.62~nm. 
We have studied their suitability to infer the longitudinal component of the magnetic field in the solar atmosphere
by applying the WFA, expanding on the work by \cite{Ishikawa2021a} and \cite{Afonso2023c}. 

We estimate that the WFA is suitable for these spectral lines if the magnetic field strength is below 1500~G. 
Therefore, the WFA should be suitable for most regions on the solar disk outside sunspots.
The exception is the \ion{Ni}{1} 280.59~nm line, for which the WFA is only suitable for magnetic field strengths below 1000~G, which means that we need to be especially careful when studying intense active regions.

The application of the WFA to the CLASP2 data confirms the suitability of these lines to infer $B_{\rm L}$ at different heights in the solar atmosphere. 
The results from the application of the WFA to all the spectral lines in the CLASP2/2.1 spectral window allowed us to estimate the formation heights of the spectral lines studied for the first time in this work.
%The analysis of the inferred values from the \ion{Ni}{1} and \ion{Fe}{2} lines allowed us to estimate that their formation height is similar to that of the \ion{Mn}{1} lines
The \ion{Ni}{1} and \ion{Fe}{2} lines seem to form at heights similar to those of the \ion{Mn}{1} lines, 
somewhere in the lower chromosphere.
When we apply the WFA to the \ion{Mn}{2} 280.62~nm line,
the inferred $B_{\rm L}$ value is larger than those inferred using any other line in the CLASP2 spectral window.
The resulting longitudinal magnetic field strength is similar to the photospheric magnetic field strength obtained from the Milne-Eddington inversion of the Hinode/SOT-SP observations of the \ion{Fe}{1} 630.15/630.25~nm lines. 
We estimate that the \ion{Mn}{2} 280.62~nm line originates in the photosphere, lower than the \ion{Fe}{2} 280.66~nm spectral line.
\cite{delPinoAleman2022} demonstrated that, although hyperfine structure effects are significant in the formation of the circular polarization profiles of the \ion{Mn}{1} resonant lines, the application of the WFA is still suitable. Future works should analyze whether this result is also applicable for the \ion{Mn}{2} 280.62~nm line, in order to confirm the B$_{\rm L}$ values inferred in this work.

We arrive to the same conclusions from the application of the WFA to the CLASP2.1 data.
In the active regions, the $B_{\rm L}$ maps inferred from the \ion{Ni}{1} and \ion{Fe}{1} lines are very similar to the maps obtained with the \ion{Mn}{1} lines and the largest $B_{\rm L}$ values are inferred from the \ion{Mn}{2} 280.62~nm line.
These results support our previous estimations.

However, the low signal-to-noise ratio in the CLASP2/2.1 circular polarization observations 
of these lines results in significant uncertainties in the inferred $B_{\rm L}$. 
Future observations with better signal-to-noise ratio will be necessary to improve the conclusions achieved in this work and to take full advantage of the magnetic diagnostic potential of these spectral lines.
Moreover, radiative transfer modeling of these lines will help us understand better their formation heights and magnetic sensitivity.

In conclusion, we confirm that the addition of the \ion{Ni}{1} 279.95 and 280.59~nm, and \ion{Fe}{1} 280.53~nm lines in the analysis of the CLASP2 and CLASP2.1 data improves and enhances the robustness of the determination of $B_{\rm L}$ in the lower layers of the solar atmosphere. 
At the same time, the addition of the \ion{Mn}{2} 280.62~nm line opens the possibility of investigating the magnetism in lower layers of the photosphere without the need for simultaneous observations with other instruments. 
This work highlights the great scientific interest of a future space telescope equipped with CLASP2-like instruments to continuously observe the solar atmosphere in this near-UV spectral window. An important example is the Chromospheric Magnetism Explorer \cite[CMEX,][]{Gilbert2022}, a NASA Small Explorer mission proposal currently undergoing Phase A evaluation. 
Following the success of our CLASP2/2.1 suborbital space experiments, such future space telescopes will be able to fully exploit the near-UV spectropolarimetry capabilities to map the stratification of the magnetic field from the photosphere to the base of the corona.

%###############################################################################
%###############################################################################
%###############################################################################
\acknowledgements

We gratefully acknowledge the financial support from the European Research Council (ERC) under the European Union’s Horizon 2020 research and innovation program (Advanced grant agreement No. 742265). 
D.A.D. is presently supported through a UCAR/ASP Postdoctoral fellowship at the 
National Center for Atmospheric Research, 
a major facility sponsored by the NSF under Cooperative Agreement No. 1852977. 
T.P.A.'s participation in the publication is part of Project RYC2021-034006-I, funded by MICIN/AEI/10.13039/501100011033, and the European Union “NextGenerationEU”/RTRP. T.d.P.A. and J.T.B. acknowledge support from the Agencia Estatal de Investigaci\'on del Ministerio de Ciencia, Innovaci\'on y Universidades (MCIU/AEI) under grant “Polarimetric Inference of Magnetic Fields” and the European Regional Development Fund (ERDF) with reference PID2022- 136563NB-I00/10.13039/501100011033. L.B. and J.T.B. gratefully acknowledge the Swiss National Science Foundation (SNSF) for financial support through grant CRSII5\_180238. 
CLASP2 and CLASP2.1 is an international partnership between NASA/MSFC, NAOJ, JAXA, IAC, and IAS; additional partners include ASCR, IRSOL, LMSAL, and the University of Oslo. 
The Japanese participation was funded by JAXA as a Small Mission-of-Opportunity Program, JSPS KAKENHI (Grant numbers JP25220703, JP16H03963, JP19K03935, and JP21K01138), 2015 ISAS Grant for Promoting International Mission Collaboration, and by 2016 NAOJ Grant for Development Collaboration.
The USA participation was funded by NASA Awards 20-HLCAS20-0010 and 16\-HTIDS16\_2\-0027. The Spanish participation was funded by the European Research Council through Advanced grant agreement No. 742265. The French hardware participation was funded by CNES funds CLASP2-13616A and 13617A.

%The Japanese participation was funded by JAXA as a Small Mission-of-Opportunity Program, 
%JSPS KAKENHI Grant numbers JP25220703 and JP16H03963, 2015 ISAS Grant for Promoting 
%International Mission Collaboration, and by 2016 NAOJ Grant for Development Collaboration. 
%The USA participation was funded by NASA Award 16-HTIDS16\_2-0027. 
%The Spanish participation was funded by the European Research Council (ERC) under the 
%European Union's Horizon 2020 research and innovation programme 
%(Advanced Grant agreement No. 742265). The French hardware participation 
%was funded by CNES funds CLASP2-13616A and 13617A.

%%%%%%%%%%%

\bibliography{feii}

\begin{thebibliography}{}
\expandafter\ifx\csname natexlab\endcsname\relax\def\natexlab#1{#1}\fi
\providecommand{\url}[1]{\href{#1}{#1}}
\providecommand{\dodoi}[1]{doi:~\href{http://doi.org/#1}{\nolinkurl{#1}}}
\providecommand{\doeprint}[1]{\href{http://ascl.net/#1}{\nolinkurl{http://ascl.net/#1}}}
\providecommand{\doarXiv}[1]{\href{https://arxiv.org/abs/#1}{\nolinkurl{https://arxiv.org/abs/#1}}}

\bibitem[{{Afonso Delgado} {et~al.}(2023){Afonso Delgado}, {del Pino Alem{\'a}n}, \& {Trujillo Bueno}}]{Afonso2023c}
{Afonso Delgado}, D., {del Pino Alem{\'a}n}, T., \& {Trujillo Bueno}, J. 2023, \apj, 954, 218, \dodoi{10.3847/1538-4357/ace4c8}

\bibitem[{{Alsina Ballester} {et~al.}(2016){Alsina Ballester}, {Belluzzi}, \& {Trujillo Bueno}}]{AlsinaBallester2016}
{Alsina Ballester}, E., {Belluzzi}, L., \& {Trujillo Bueno}, J. 2016, \apjl, 831, L15, \dodoi{10.3847/2041-8205/831/2/L15}

\bibitem[{{Belluzzi} \& {Trujillo Bueno}(2012)}]{Belluzzi-TrujilloBueno2012}
{Belluzzi}, L., \& {Trujillo Bueno}, J. 2012, \apjl, 750, L11, \dodoi{10.1088/2041-8205/750/1/L11}

\bibitem[{{del Pino Alem{\'a}n} {et~al.}(2022){del Pino Alem{\'a}n}, {Alsina Ballester}, \& {Trujillo Bueno}}]{delPinoAleman2022}
{del Pino Alem{\'a}n}, T., {Alsina Ballester}, E., \& {Trujillo Bueno}, J. 2022, \apj, 940, 78, \dodoi{10.3847/1538-4357/ac922c}

\bibitem[{{del Pino Alem{\'{a}}n} {et~al.}(2016){del Pino Alem{\'{a}}n}, Casini, \& {Manso Sainz}}]{DelPinoAleman2016}
{del Pino Alem{\'{a}}n}, T., Casini, R., \& {Manso Sainz}, R. 2016, The Astrophysical Journal, 830, L24, \dodoi{10.3847/2041-8205/830/2/l24}

\bibitem[{{del Pino Alem{\'{a}}n} {et~al.}(2020){del Pino Alem{\'{a}}n}, {Trujillo Bueno}, Casini, \& {Manso Sainz}}]{delPinoAleman2020}
{del Pino Alem{\'{a}}n}, T., {Trujillo Bueno}, J., Casini, R., \& {Manso Sainz}, R. 2020, The Astrophysical Journal, 891, 91, \dodoi{10.3847/1538-4357/ab6bc9}

\bibitem[{{Fontenla} {et~al.}(1993){Fontenla}, {Avrett}, \& {Loeser}}]{Fontenla1993}
{Fontenla}, J.~M., {Avrett}, E.~H., \& {Loeser}, R. 1993, \apj, 406, 319, \dodoi{10.1086/172443}

\bibitem[{{Gilbert}(2022)}]{Gilbert2022}
{Gilbert}, H.~R. 2022, in AGU Fall Meeting Abstracts, Vol. 2022, SH25D--2079

\bibitem[{Ishikawa {et~al.}(2021)Ishikawa, Trujillo~Bueno, {Del Pino Aleman}, Okamoto, McKenzie, Auchere, Kano, Song, Yoshida, Rachmeler, Kobayashi, Hara, Kubo, Narukage, Sakao, Shimizu, Suematsu, Bethge, {De Pontieu}, Dalda, Vigil, Winebarger, Ballester, Belluzzi, {\v{S}}tepan, Ramos, Carlsson, \& Leenaarts}]{Ishikawa2021a}
Ishikawa, R., Trujillo~Bueno, J., {Del Pino Aleman}, T., {et~al.} 2021, Science Advances, 7, 1, \dodoi{10.1126/sciadv.abe8406}

\bibitem[{Kramida {et~al.}(2020)Kramida, {Yu.~Ralchenko}, Reader, \& {and NIST ASD Team}}]{NIST_ASD}
Kramida, A., {Yu.~Ralchenko}, Reader, J., \& {and NIST ASD Team}. 2020, {NIST Atomic Spectra Database (ver. 5.8), [Online]. Available: {\tt{https://physics.nist.gov/asd}} [2017, April 9]. National Institute of Standards and Technology, Gaithersburg, MD.}

\bibitem[{{Landi Degl'Innocenti} \& {Landolfi}(2004)}]{LL04}
{Landi Degl'Innocenti}, E., \& {Landolfi}, M. 2004, {Polarization in Spectral Lines}, Vol. 307 (Dordrecht: Sprintger), \dodoi{10.1007/978-1-4020-2415-3}

\bibitem[{{Li} {et~al.}(2022){Li}, {del Pino Alem{\'a}n}, {Trujillo Bueno}, \& {Casini}}]{Li2022}
{Li}, H., {del Pino Alem{\'a}n}, T., {Trujillo Bueno}, J., \& {Casini}, R. 2022, \apj, 933, 145, \dodoi{10.3847/1538-4357/ac745c}

\bibitem[{{Li} {et~al.}(2023){Li}, {del Pino Alem{\'a}n}, {Trujillo Bueno}, {Ishikawa}, {Alsina Ballester}, {McKenzie}, {Auch{\`e}re}, {Kobayashi}, {Okamoto}, {Rachmeler}, \& {Song}}]{Li2023}
{Li}, H., {del Pino Alem{\'a}n}, T., {Trujillo Bueno}, J., {et~al.} 2023, \apj, 945, 144, \dodoi{10.3847/1538-4357/acb76e}

\bibitem[{{Li} {et~al.}(2024){Li}, {del Pino Alem{\'a}n}, {Trujillo Bueno}, {Ishikawa}, {Alsina Ballester}, {McKenzie}, {Belluzzi}, {Song}, {Okamoto}, {Kobayashi}, {Rachmeler}, {Bethge}, \& {Auch{\`e}re}}]{Li2024}
---. 2024, arXiv e-prints, arXiv:2408.06094, \dodoi{10.48550/arXiv.2408.06094}

\bibitem[{{Mart{\'\i}nez Gonz{\'a}lez} {et~al.}(2012){Mart{\'\i}nez Gonz{\'a}lez}, {Manso Sainz}, {Asensio Ramos}, \& {Belluzzi}}]{MartinezGonzalez2012}
{Mart{\'\i}nez Gonz{\'a}lez}, M.~J., {Manso Sainz}, R., {Asensio Ramos}, A., \& {Belluzzi}, L. 2012, \mnras, 419, 153, \dodoi{10.1111/j.1365-2966.2011.19681.x}

\bibitem[{{Narukage} {et~al.}(2016){Narukage}, {McKenzie}, {Ishikawa}, {Trujillo-Bueno}, {De Pontieu}, {Kubo}, {Ishikawa}, {Kano}, {Suematsu}, {Yoshida}, {Rachmeler}, {Kobayashi}, {Cirtain}, {Winebarger}, {Asensio Ramos}, {del Pino Aleman}, {{\v{S}}t{\k{e}}p{\'a}n}, {Belluzzi}, {Larruquert}, {Auch{\`e}re}, {Leenaarts}, \& {Carlsson}}]{Narukage2016}
{Narukage}, N., {McKenzie}, D.~E., {Ishikawa}, R., {et~al.} 2016, in Society of Photo-Optical Instrumentation Engineers (SPIE) Conference Series, Vol. 9905, Space Telescopes and Instrumentation 2016: Ultraviolet to Gamma Ray, ed. J.-W.~A. {den Herder}, T.~{Takahashi}, \& M.~{Bautz}, 990508, \dodoi{10.1117/12.2232245}

\bibitem[{{Song} {et~al.}(2018){Song}, {Ishikawa}, {Kano}, {Yoshida}, {Tsuzuki}, {Uraguchi}, {Shinoda}, {Hara}, {Okamoto}, {Auch{\`e}re}, {McKenzie}, {Rachmeler}, \& {Trujillo Bueno}}]{Song2018}
{Song}, D., {Ishikawa}, R., {Kano}, R., {et~al.} 2018, in Society of Photo-Optical Instrumentation Engineers (SPIE) Conference Series, Vol. 10699, Space Telescopes and Instrumentation 2018: Ultraviolet to Gamma Ray, ed. J.-W.~A. {den Herder}, S.~{Nikzad}, \& K.~{Nakazawa}, 106992W, \dodoi{10.1117/12.2313056}

\bibitem[{{Song} {et~al.}(2022){Song}, {Ishikawa}, {Kano}, {McKenzie}, {Trujillo Bueno}, {Auch{\`e}re}, {Rachmeler}, {Okamoto}, {Yoshida}, {Kobayashi}, {Bethge}, {Hara}, {Shinoda}, {Shimizu}, {Suematsu}, {De Pontieu}, {Winebarger}, {Narukage}, {Kubo}, {Sakao}, {Asensio Ramos}, {Belluzzi}, {{\v{S}}t{\v{e}}p{\'a}n}, {Carlsson}, {del Pino Alem{\'a}n}, {Alsina Ballester}, {Vigil}, \& {Leenaarts}}]{Song2022}
{Song}, D., {Ishikawa}, R., {Kano}, R., {et~al.} 2022, \solphys, 297, 135, \dodoi{10.1007/s11207-022-02064-8}

\bibitem[{{Trujillo Bueno} \& {del Pino Alem{\'a}n}(2022)}]{TrujilloBueno2022}
{Trujillo Bueno}, J., \& {del Pino Alem{\'a}n}, T. 2022, \araa, 60, 415, \dodoi{10.1146/annurev-astro-041122-031043}

\bibitem[{{Tsuzuki} {et~al.}(2020){Tsuzuki}, {Ishikawa}, {Kano}, {Narukage}, {Song}, {Yoshida}, {Uraguchi}, {Okamoto}, {McKenzie}, {Kobayashi}, {Rachmeler}, {Auchere}, \& {Trujillo Bueno}}]{Tsuzuki2020}
{Tsuzuki}, T., {Ishikawa}, R., {Kano}, R., {et~al.} 2020, in Society of Photo-Optical Instrumentation Engineers (SPIE) Conference Series, Vol. 11444, Space Telescopes and Instrumentation 2020: Ultraviolet to Gamma Ray, ed. J.-W.~A. {den Herder}, S.~{Nikzad}, \& K.~{Nakazawa}, 114446W, \dodoi{10.1117/12.2562273}

\bibitem[{{Yoshida} {et~al.}(2018){Yoshida}, {Song}, {Ishikawa}, {Kano}, {Katsukawa}, {Suematsu}, {Narukage}, {Kubo}, {Shinoda}, {Okamoto}, {McKenzie}, {Rachmeler}, {Auch{\`e}re}, \& {Trujillo Bueno}}]{Yoshida2018}
{Yoshida}, M., {Song}, D., {Ishikawa}, R., {et~al.} 2018, in Society of Photo-Optical Instrumentation Engineers (SPIE) Conference Series, Vol. 10699, Space Telescopes and Instrumentation 2018: Ultraviolet to Gamma Ray, ed. J.-W.~A. {den Herder}, S.~{Nikzad}, \& K.~{Nakazawa}, 1069930, \dodoi{10.1117/12.2312463}

\end{thebibliography}
\bibliographystyle{aasjournal}

\appendix 
\section{Spectral degradation} \label{sec:appA}
\begin{figure*}[htp!]
    \centering
    \includegraphics[width=.4\textwidth]{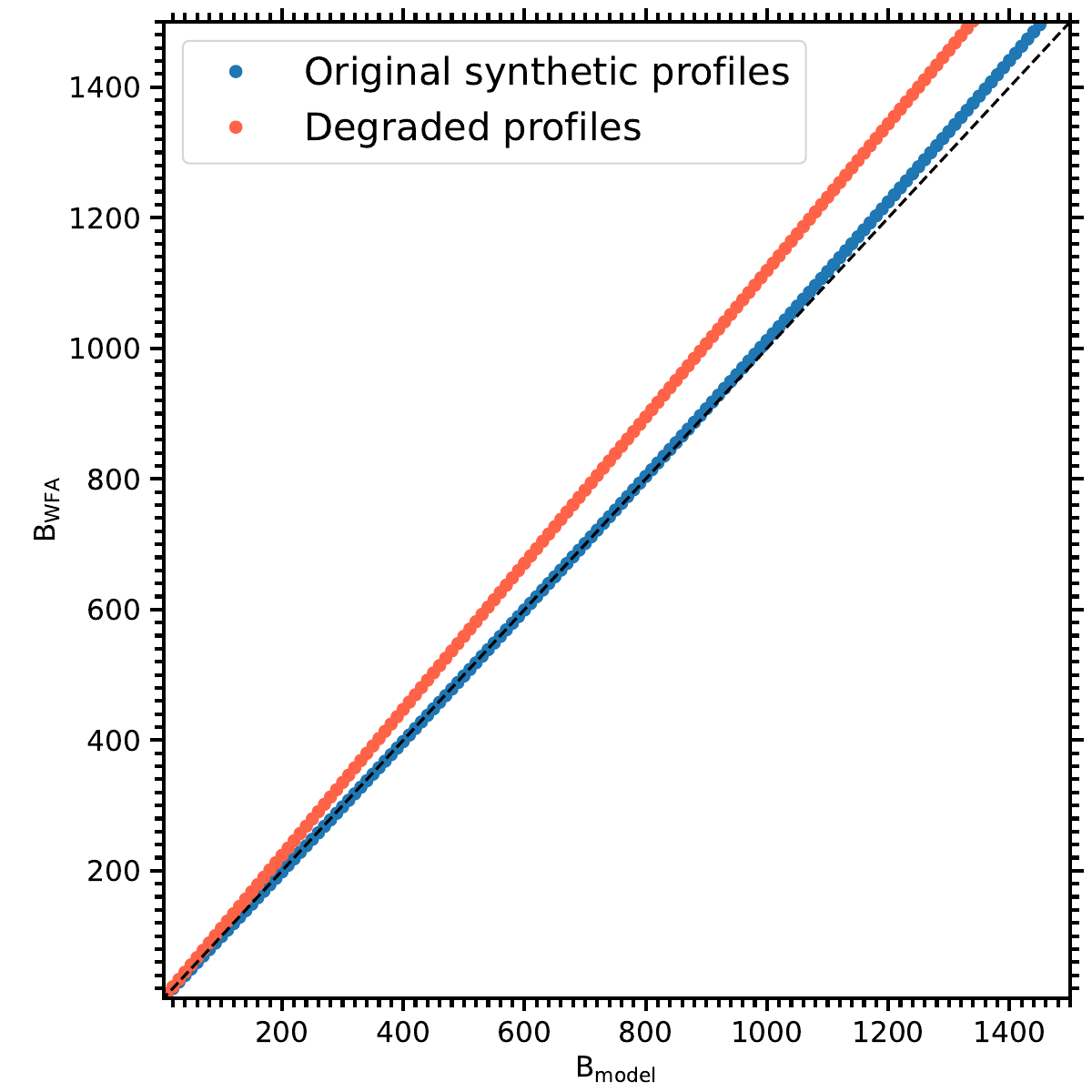}
    \caption{Inferred B$_{\rm L}$ vs the original value from the application of the WFA to the full resolution profiles (blue) and to the profiles degraded to the CLASP2 spectral resolution (red).}
    \label{fig:specDgr}
\end{figure*}

Figure~\ref{fig:specDgr} shows the longitudinal magnetic field inferred from the application of the WFA to synthetic Stokes profiles of the \ion{Mg}{2} h and k spectral lines, before applying any spectral degradation (blue) and after degrading the profiles to the CLASP2 spectral resolution (red). The profiles were computed in a static FAL-C atmosphere with a uniform longitudinal magnetic field from 10 to 1500~G. The spectral degradation introduces a systematic overestimation of around 15~\%.

\end{document}